\documentstyle[12pt]{article}

\newcommand{\rA}{\rightarrow}
\newcommand{\ti}{\rightarrow \infty}
\newcommand{\tz}{\rightarrow 0}
\newtheorem{theorem}{Theorem}
\newcommand{\eb}{\makebox}

\begin{document}

\title{On limits of spacetimes -- a coordinate-free approach}
\author{
F. M. Paiva\thanks{
{\sc internet: fmpaiva@cbpfsu1.cat.cbpf.br;\ \ bitnet: fmpaiva@brlncc}},\ \
M. J. Rebou\c{c}as\thanks{{\sc bitnet: reboucas@brlncc}}\ \ and\ \
M. A. H. MacCallum\thanks{{\sc internet: M.A.H.MacCallum@maths.qmw.ac.uk}} \\
\\
$^{\ast}~^{\dagger}~$Centro Brasileiro de Pesquisas F\'\i sicas, \\
R. Dr. Xavier Sigaud 150, \\
22290-180 Rio de Janeiro -- RJ, Brazil \\
\\
$^{\ddagger}~$School of Mathematical Sciences, \\
Queen Mary and Westfield College, \\
Mile End Road,
London E1 4NS -- U. K.
}
\date{}
\maketitle
\begin{abstract}
A coordinate-free approach to limits of spacetimes is developed. The limits of
the Schwarzschild metric as the mass parameter tends to $0$ or $\infty$ are
studied, extending previous results. Besides the known Petrov type D and 0
limits, three vacuum plane-wave solutions of Petrov type N are found to be
limits of the Schwarzschild spacetime.
\end{abstract}
{\sc pacs} numbers: 04.20.Jb, 04.20.Cv

\section{Introduction} \label{int} \setcounter{equation}{0}

Many calculations in general relativity are done in a specific coordinate
system. However, distinct results might arise if different coordinate systems
were used. Concerning limits, Geroch \cite{Geroch} showed that the
Schwarzschild spacetime may have as its limit Minkowski space, a Kasner
spacetime or even a singular limit as the mass tends to infinity, depending on
the coordinate system in which the limit is taken. Actually, by limit of a
spacetime we mean limit of a family of spacetimes as some free parameters are
taken to a limit.  For instance, in the one-parameter family of Schwarzschild
solutions each member is a Schwarzschild spacetime with a specific value for
the mass parameter $m$.

In this paper, using a coordinate-free characterization of metrics
borrowed from the equivalence problem \cite{Cartan,Karlhede}, we
develop an approach to finding the limits of a given
spacetime\footnote{Although we have focused our attention on a Petrov
type D vacuum spacetime, the coordinate-free approach devised in this
paper can be extended to non-vacuum spacetimes with different Petrov
types.}. Also, we apply our method to the Schwarzschild metric,
recovering and extending the results in the literature \cite{Geroch}.

In section~\ref{eq} we fix the notation, introduce the equivalence problem and
show how it leads to a coordinate-free characterization of spacetimes and,
therefore, to a coordinate-free approach to limits. In section~\ref{s} we
apply this technique to the study of the limits of the Schwarzschild solution.
Section \ref{con} concludes with some considerations about further
improvements of the method and its application to other situations.

\section{The coordinate-free approach} \label{eq} \setcounter{equation}{0}

In this section, the equivalence problem and its solution due to Cartan
\cite{Cartan} and Karlhede \cite{Karlhede} are presented. We show that it
leads to a coordinate-free description of spacetimes, which is the basis of our
approach to dealing with limits. The main features of this new method are
discussed --- the details appear, as it is applied to the Schwarzschild
spacetime, in the next section.

Given two spacetimes $M$ and $\tilde{M}$, in general relativity, one might
wish to decide whether or not they are locally equivalent (the equivalence
problem), i.e., whether a coordinate transformation
\begin{equation}
\label{cord}\tilde{x}^a = \tilde{x}^a(x)
\end{equation}
exists such that
\begin{equation}
\label{met}
g_{ab}(x) = \frac{\partial \tilde{x}^c}{\partial x^a}
            \frac{\partial \tilde{x}^d}{\partial x^b}
            \tilde{g}_{cd}(\tilde{x})
\end{equation}
holds, where $g_{ab}(x)$ and $\tilde{g}_{cd}(\tilde{x})$ are the components of
the metric tensor of each spacetime in the coordinates $x$ and $\tilde{x}$
respectively. A first attempt to solve this problem would probably be to use
the scalars made by contraction of the Riemann tensor and its covariant
derivatives \cite{ISCA}. Unfortunately this fails since, for example, all
these scalars vanish for plane-wave and Minkowski spacetimes and yet they are
not the same.  The best known solution to the equivalence problem was
presented by Cartan \cite{Cartan} and can be summarized as follows.

Let $\omega^A~(A = 0,\ldots,3)$ be a frame such that the line element can be
written as
\begin{equation}
ds^2 = \eta_{AB} ~\omega^A ~\omega^B . \label{mink}
\end{equation}
In a constant frame ($\eta_{AB}$ constant), find the components of the Riemann
tensor and its covariant derivatives up to (possibly) the $10^{\rm th}$ order
for both metrics. Then:
\begin{theorem}\label{eqt}The two metrics are equivalent if and only if
there exist coordinate and Lorentz transformations which transform one set
of components into the other.
\end{theorem}

Notice that there may be 11 steps, from the $0^{\rm th}$ order derivative up
to the $10^{\rm th}$. This number is somehow related to the 6 Lorentz
transformations, the 4 coordinates of the spacetime and one integrability
condition. Karlhede \cite{Karlhede} introduced a procedure, used in practice,
in which the coordinate and Lorentz transformations are treated differently. A
fixed frame is chosen to perform the calculations so that only the coordinates
appear in the components of the Riemann tensor and not the Lorentz parameters.
To describe the Karlhede algorithm, we introduce the following notation:
\begin{description}
\item $R_{n}$: the set of components of the covariant derivatives of the
Riemann tensor from the $0^{\rm th}$ to the $n^{\rm th}$ order --- the {\em
Cartan scalars};
\item $C_{n}$: the algebraically independent Cartan scalars among $R_{n}$;
\item $G_{n}$: the isometry group of $R_{n}$;
\item $H_{n}$: the isotropy group of $R_{n}$;
\item $t_{n}$: the number of independent functions of the coordinates in
$R_{n}$;
\item $d_{n}$: the dimension of the orbit of $G_{n}$, given by $d_{n} =
4 - t_{n}$. Note that ${\rm dim}(G_{n}) = d_{n} + {\rm dim}(H_{n})$.
\end{description}

The $0^{\rm th}$ step of the algorithm is the following. Choose a tetrad basis
in which the metric is constant. Calculate $R_{0}$ and, using Lorentz
transformations, find a basis where it takes a well-defined standard form
\cite{Karlhede}. Then find $H_{0}$ and $d_{0}$. $H_{0}$ is found with the help
of the Petrov and Segre classifications, while $d_{0}$ is the dimension of the
spacetime minus $t_{0}$. The next $n$ steps are similar to each
other, as described below.

Calculate $R_{n}$ and, using $H_{n-1}$, find a basis where $R_{n}$ takes a
well-defined standard form \cite{Karlhede}. Then find $H_{n}$ and $d_{n}$.
Now, $H_{n}$ is easily found as it must be a subgroup of $H_{n-1}$, which is
usually of low dimension. If $H_{n} \subset H_{n-1}$ (strictly, i.e.\
$H_{n} \neq H_{n-1}$) or $d_{n} < d_{n-1}$, the
algorithm continues for another step. If $H_{n} = H_{n-1}$ and $d_{n} =
d_{n-1}$, then $R_{n}$ can be expressed as functions of $R_{n-1}$, and so the
algorithm stops. In most cases the algorithm stops at the $2^{\rm nd}$ or
$3^{\rm rd}$ covariant derivative --- the $4^{\rm th}$ is the highest
derivative necessary until now \cite{Koutras}.

The set $R_{n}$ must be calculated in this way for the two metrics
individually. As the frames are already fixed in both cases, we have only to
verify whether or not coordinate transformations exist which transform one set
of components to the other as stated in theorem \ref{eqt}. Although this step
is not algorithmic, in most cases it can be tackled with the help of a
computer algebra system. All the algorithmic steps are already implemented in
the computer algebra system {\sc sheep} \cite{Frick}, and constitute a suite
of computer algebra programs called {\sc classi} \cite{Aman}. A database of
spacetimes together with their Cartan scalars has been produced using {\sc
classi} \cite{ISCA}.

Not all components of the Riemann tensor and its covariant derivatives are
algebraically independent; working with spinors MacCallum and {\AA}man
\cite{AmanMac} found a minimal set of algebraically independent quantities
which are actually calculated in the algorithm.  Their components will be
called {\em algebraically independent Cartan scalars} and denoted by $C_n$,
according to the order of derivation. For vacuum solutions, as is the case of
Schwarzschild, $C_2$ is:
\begin{eqnarray}
0^{\rm th}~{\rm derivative}:& & \Psi_A,~A = (0,\ldots,4);
\label{c0} \\
1^{\rm st}~{\rm derivative}:& & D\Psi_{A B^{'}},~A = (0,\ldots,5),~B = (0,~1);
\label{c1} \\
2^{\rm nd}~{\rm derivative}:& & D^2\Psi_{A B^{'}},~A = (0,\ldots,6),~B =
(0,~1,~2),
\label{c2}
\end{eqnarray}
where $\Psi_A$ is the Weyl spinor and $D\Psi_{A B^{'}}$ and $D^2\Psi_{A B{'}}$
are the first and second symmetrized covariant derivatives of $\Psi_A$,
respectively \cite{ISCA}.

The Cartan scalars are scalars under coordinate transformations. Also, they
are sufficient to decide the equivalence of metrics, so they contain all local
information about a metric and provide a coordinate-free local
characterization of a spacetime. Nevertheless, for each frame used, a
different set of Cartan scalars may be obtained. Although in the equivalence
problem this has been overcome by choosing standard forms for the scalars
which fix the frame up to isometries of the spacetime itself, in what concerns
characterizations of metrics they are not {\em frame-free}. In other words,
they are scalars on the frame bundle and become scalars on the manifold only
after a choice of a canonical frame.

It could appear at first sight that the coordinate arbitrariness is just
substituted by the frame freedom, but this is not so. Coordinates transform
according to the manifold mapping group (MMG), with infinite dimension.
Working with components in a coordinate basis, the group at each point is
GL(4,R), with dimension 16.  Frames also transform according to GL(4,R), but
one can always impose a specific metric $\eta_{AB}$ in (\ref{mink}), Minkowski
for instance, and the frame transformations become restricted to the Lorentz
group, with 6 dimensions.

Therefore, instead of performing limits on the metric tensor subjected to
GL(4,R), we shall work with scalars, the Cartan scalars, and the Lorentz
group. Taking into account information given by the Petrov (and Segre for
non-vacuum spacetimes) classification, as we shall discuss in the next
section, only subgroups of the Lorentz group will have to be considered.
Classifications of the covariant derivatives of the Riemann tensor would also
be very helpful, but unfortunately they do not yet exist.

In order to use the Cartan scalars and the Petrov and Segre classifications in
the study of limits, we shall take into account one of Geroch's \cite{Geroch}
hereditary properties of limits. He shows that if a tensor, vector or scalar
constructed from the metric and its derivatives vanishes for all members of a
family of spacetimes, it must also vanish for any limit of this family. So, he
concludes, limits of vacuum spacetimes ($R_{ab} = 0$) are vacuum, and limits
of a Petrov type are either of the same Petrov type or one of its
specializations. This follows from the fact that some scalars constructed from
the Weyl tensor vanish as we go down through its specializations.

The notation used for the Lorentz transformations and their effects on the
Riemann spinor may be found in \cite{Karlhede}. The effect on the covariant
derivatives of the Riemann spinor is just a generalization of this result to
higher order symmetrized spinors.

\section{Limits of the Schwarzschild spacetime} \label{s}
\setcounter{equation}{0}

The Schwarzschild line element may be written as
\begin{equation}
ds^{2} = A dt^{2} - A^{-1} dr^{2}
         - r^{2}(d\theta ^{2} + \sin^{2}\theta d\phi ^{2}),  \label{s1}
\end{equation}
where
\begin{equation}
A = 1 - \frac{2m}{r}.  \label{A}
\end{equation}
In this coordinate system the limit of the component $g_{00}$ of the metric
diverges as $m\ti$ . Nevertheless,
Geroch \cite{Geroch} shows that in the coordinate systems defined by
\begin{equation}
\tilde{r}=m^{-1/3} r,~~\tilde{t}=m^{1/3} t,~~\rho=m^{1/3} \theta \label{s2}
\end{equation}
and
\begin{equation}
x = r + m^{4/3}, ~~ \rho = m^{4/3} \theta,  \label{s3}
\end{equation}
the limits of the metric are a Kasner spacetime and Minkowski space
respectively. Following the reasoning outlined in the previous section, we
shall now calculate the Cartan scalars and study their limits.

In the null tetrad
\begin{eqnarray}
\omega^{0} = \frac{1}{\sqrt{2}}(\theta^{0} + \theta^{1}),~~
\omega^{1} = \frac{1}{\sqrt{2}}(\theta^{0} - \theta^{1}),~ \nonumber \\
\label{nullt} \\
\omega^{2} = \frac{1}{\sqrt{2}}(\theta^{2} + i \theta^{3}),~~
\omega^{3} = \frac{1}{\sqrt{2}}(\theta^{2} - i \theta^{3}),   \nonumber
\end{eqnarray}
where $\theta^{A}$ is a Lorentz tetrad
($\eta_{AB} = {\rm diag}\,(+1,-1,-1,-1)$) given by
\begin{equation}
\theta^{0} = A^{1/2}       dt,~~
\theta^{1} = A^{-1/2}      dr,~~
\theta^{2} = r             d\theta,~~
\theta^{3} = r \sin\theta d\phi,          \label{ts1}
\end{equation}
the nonzero algebraically independent Cartan scalars ($C_n$) are:
\begin{eqnarray}
\eb[12em][r]{$0^{\rm th}$ derivative:\hfill
$\Psi_{2}$}
& = & -\frac{m}{r^3}, \label{i0} \\
& & \nonumber \\
\eb[12em][r]{$1^{\rm st}$ derivative:\hfill
$D\Psi_{2 0^{'}}$} & = &
 ~~\frac{3}{\sqrt{2}} \frac{m}{r^4} \sqrt{1 - \frac{2m}{r}}, \label{i1a} \\
\eb[12em][r]{$D\Psi_{3 1^{'}}$} & = &
-\frac{3}{\sqrt{2}} \frac{m}{r^4} \sqrt{1 - \frac{2m}{r}}, \label{i1b} \\
& & \nonumber \\
\eb[12em][r]{$2^{\rm nd}$ derivative:\hfill
$D^2\Psi_{2 0^{'}}$}& = & \frac{12m^2}{r^6} - \frac{6m}{r^5},  \label{i2a} \\
\eb[12em][r]{$D^2\Psi_{3 1^{'}}$} & = &
\frac{27m^2}{2r^6} - \frac{6m}{r^5}, \label{i2b} \\
\eb[12em][r]{$D^2\Psi_{4 2^{'}}$} & = &
\frac{12m^2}{r^6} - \frac{6m}{r^5}. \label{i2c}
\end{eqnarray}
Some useful information such as Petrov and Segre types and dimension of the
isometry and isotropy groups can be found directly from the Cartan scalars.
The easiest way of doing this is to follow the steps of the Karlhede
algorithm, as we do below.

The only nonzero member of $C_0$ is $\Psi_2$; the Ricci spinor and the
curvature scalar vanish. So we are dealing with a vacuum solution, as
expected, and from now on only covariant derivatives of the Weyl spinor will
appear. The Petrov type is D so the standard form for $\Psi_A$ is the above
one (eq. (\ref{i0})) \cite{Karlhede}. $H_{0}$ for this Petrov type in this
canonical frame is the 2-parameter group of boosts in the
$\omega^{0}-\omega^{1}$ plane
\begin{equation} \label{boost} \left( \begin{array}{cc}
a & 0 \\
0 & a^{-1} \\
\end{array} \right), \end{equation}
where $a$ is real, and spatial rotations in the $\omega^{2}-\omega^{3}$ plane
\begin{equation} \label{spin} \left( \begin{array}{cc}
e^{i\theta} & 0 \\
0           & e^{- i\theta} \\
\end{array} \right), \end{equation}
where $\theta$ is real \cite{Karlhede}. There is only one independent function
of the coordinates ($\Psi_{2}$), so $t_{0} = 1$ and $d_{0} = 3$.

In the next step of the algorithm, we calculate $C_{1}$ which contains $C_{0}$
with $D\Psi_{2 0^{'}}$ and $D\Psi_{3 1^{'}}$. Under $H_{0}$ the nonzero
components of the transformed $D\Psi_{A B^{'}}$ are
\begin{eqnarray}
\widetilde{D\Psi}_{2 0^{'}} & = & a^{2} D\Psi_{2 0^{'}}~, \label{boosti1a} \\
\widetilde{D\Psi}_{3 1^{'}} & = & a^{-2} D\Psi_{3 1^{'}}~. \label{boosti1b}
\end{eqnarray}
The parameter $a$ can be chosen in order to make these two components equal
up to a sign. As the spatial rotation does not affect them, $H_{1}$ is the
1-parameter group of spatial rotations (\ref{spin}). There is no new
independent function, so $d_{1} = 3$.

We proceed with the algorithm and find $C_{2}$ as expressions
(\ref{i0})--(\ref{i2c}). They can be shown to be invariant under the spatial
rotations (\ref{spin}) and also they contain no new independent function.  So
$H_{2} = H_{1}$ and $d_{2} = d_{1}$ and the algorithm stops. Therefore the
orbit of the isometry group has dimension 3 and the isotropy group is
one-dimensional. So the isometry group has dimension 4. Having this
information, we can now begin the study of the limits of the Schwarzschild
family.

If the limit as $m\ti$ is taken, using the Cartan scalars given by
(\ref{i0})--(\ref{i2c}), the result is a divergent limit. Nevertheless, if the
coordinate transformations (\ref{s2}) and (\ref{s3}) are performed on these
scalars, regular limits are obtained and the resulting Cartan scalars can be
shown to be those of a Kasner spacetime and Minkowski space, respectively.
So, although we are explicitly imposing a limit only on the parameter $m$, a
choice of a coordinate system implicitly imposes a limit on the Cartan scalars
which is outwith our control. This means that a limit of a spacetime is not
entirely defined just by the limit of the mass $m$~---~we have to say
explicitly what are the limits of all Cartan scalars in order to have a single
limit spacetime. This choice of the limits of the Cartan scalars is not,
however, completely arbitrary as there are relations between the Cartan
scalars derived from (\ref{i0})--(\ref{i2c}) which are hereditary in the sense
of Geroch. Indeed, eliminating $r$ from (\ref{i1a})--(\ref{i2c}), the Cartan
scalars for the Schwarzschild spacetime can be written\footnote{Incidentally,
eqs.~(\ref{i0}) and (\ref{i1a}) may be solved to give $r$ and $m$ as functions
of $\Psi_2$ and $D\Psi_{2 0^{'}}$, so a mass and a radial distance can be
defined in a coordinate-free way.} as:
\begin{eqnarray}
\eb[12em][r]{$0^{\rm th}$ derivative:\hfill
$\Psi_{2}$}
& = &  -\frac{m}{r^3}, \label{r0} \\
& & \nonumber \\
\eb[12em][r]{$1^{\rm st}$ derivative:\hfill
$D\Psi_{2 0^{'}}$}
& = & \frac{3}{\sqrt{2}}\Psi_2^{\frac{4}{3}}
\sqrt{m^{-\frac{2}{3}} + 2\Psi_2^{\frac{1}{3}}}, \label{r1a} \\
\eb[12em][r]{$D\Psi_{3 1^{'}}$} & = & - D\Psi_{2 0^{'}}, \label{r1b} \\
& & \nonumber \\
\eb[12em][r]{$2^{\rm nd}$ derivative:\hfill
$D^2\Psi_{2 0^{'}}$} & = & \frac{4}{3} \frac{(D\Psi_{2 0^{'}})^2}{\Psi_2},
\label{r2a} \\
\eb[12em][r]{$D^2\Psi_{3 1^{'}}$}
& = & - D^2\Psi_{2 0^{'}}-\frac{3}{2} \Psi_2^2~, \label{r2b} \\
\eb[12em][r]{$D^2\Psi_{4 2^{'}}$} & = & D^2\Psi_{2 0^{'}}~. \label{r2c}
\end{eqnarray}
Expressions (\ref{r1a})--(\ref{r2c}) are coordinate-free so they hold for any
limit of the Schwarzschild spacetime. Therefore, once limits of $m$ and
$\Psi_{2}$ are chosen, the limits of all the other scalars have to satisfy
these expressions.

Eq.~(\ref{r0}) is valid in a specific coordinate system and, with suitable
coordinate transformations, its dependence on $m$ can be made arbitrary.
Therefore, the limits of $\Psi_{2}$ and $m$ can be chosen independently. We
shall study the limits as the mass tends to $0$ or $\infty$ combined with the
four possible limits for $\Psi_{2}$, namely, (a) 0; (b) a function
of the four coordinates, $f(x^i)$; (c) a non-zero constant; or (d)
$\infty$.

Before imposing the limit on $m$, it is worth checking whether the above
limits satisfy basic integrability conditions, i.e., whether a metric could
exist with those values for $\Psi_{A}$. Case (a) is Minkowski space, provided
that all covariant derivatives vanish, and (b) is possible as shown below. The
case (c) is not possible as the Newman-Penrose equations \cite{KSMH} become
incompatible if we impose vacuum together with the condition $\Psi_{2} =
const$ and $\Psi_{A} = 0,~ A \neq 2$. Finally, although (d) is possible, in
this paper we shall not consider singular limits, i.e., limits where a scalar
tends to $\infty$ in some region of the spacetime. To study these limits
junction conditions would have to be analysed and this is beyond the scope of
the present paper.

We shall now discuss the two remaining possibilities, namely (a) and
(b), for each choice of the limit of $m$.
\begin{enumerate}
\item $m\ti$ \begin{enumerate}
 \item $\Psi_{2}\tz$. From (\ref{r1a})--(\ref{r2c}) all Cartan scalars tend
to zero. So Minkowski space is a limit of the Schwarzschild spacetime.
 \item $\Psi_{2}\rA f(x^i)$. From (\ref{r1a}), $D\Psi_{2 0^{'}}\rA
3(f(x^i))^{3/2}$. In principle this could lead to many different limit
spacetimes, depending on the function $f(x^i)$.  Theorem \ref{eqt} can then be
used to check whether those spacetimes are really inequivalent for different
functions, say $f(x^i)$ and $g(\tilde x^i)$. As eqs.~(\ref{r1a})--(\ref{r2c})
hold for any limit, only the compatibility of
\begin{equation} \label{eqtest}
f(x^i) = g(\tilde x^i)
\end{equation}
is left to be checked. However, this equation just defines the coordinate
transformations necessary to transform one set of Cartan scalars into the
other.
Therefore, this case consists of a single limit.
 \end{enumerate}
\item $m\tz$ \begin{enumerate}
 \item $\Psi_{2}\tz$. Similarly to (1.a), this limit is Minkowski
space.
 \item $\Psi_{2}\rA f(x^i)$. From (\ref{r1a}), $D\Psi_{2 0^{'}}\ti$.
Therefore, this is a singular limit.
 \end{enumerate}
\end{enumerate}

The integrability conditions for the full set of scalars obtained would have
to be checked now. This is in general a difficult task
\cite{KarlhedeLindstron}. Nevertheless, cases (1.a) and (2.a)
are easily recognized as Minkowski space and the metric of case (1.b)
is known to be a Kasner metric \cite{Geroch}.

One could now ask what are the three coordinate systems in which the
Schwarzschild metric has the limits above --- do they exist? We still do not
know how to answer such questions in the general case; nevertheless for
Schwarzschild we can present the coordinate systems. They are those given by
Geroch \cite{Geroch} for the two cases where $m \ti$ and the one given in
(\ref{s1}) for $m \tz$.

Now we proceed with the analysis of the Lorentz freedom of the definition of
the frame, referred to in section~\ref{eq}.  From Geroch's hereditary property
we conclude that Petrov type D solutions may have as a limit only Petrov types
D, N or 0 \cite{Penrose}. Nevertheless, in the standard frame we chose, only
$\Psi_2$ is different from zero in $C_0$ and it can be shown \cite{Karlhede}
that this condition is impossible for Petrov type N metrics. Moreover, from
the hereditary property, scalars which vanish for all members of a family must
also be zero in the limit, so with the Cartan scalars calculated in this frame
it will not be possible to study Petrov type N limits (Petrov type 0 limits
present no problem as all components of the $\Psi_{A}$ vanish identically).
Therefore, it is necessary to transform to another frame where the Cartan
scalars may have a Petrov type N limit. The parameters of this transformation
ought to be defined in a coordinate-free way --- they should be functions of
the Cartan scalars.

As the analysis in the first frame seems to have exhausted all possible limits
of the Schwarzschild spacetime, it looks as though no new limit can
arise now.  Nevertheless, it was assumed that the limit as $\Psi_2 \tz$ had to
be Minkowski space, and this is not necessary now. Indeed, if we choose an
infinite limit for the frame transformation parameters, the limit of the
transformed $C_0$ might not be zero as $\Psi_2 \tz$ and might be Petrov type
N. One may question how it can be a Petrov type N solution if all Cartan
scalars vanish in the first frame (this should be Minkowski space!).  The
answer to this question seems to be that before the limit is taken, the two
frames are related by a Lorentz transformation, therefore they describe the
same spacetime. But after taking the limit, the frames are related by an
``infinite Lorentz transformation'', which actually is not a Lorentz
transformation. Therefore, these two frames correspond to different
spacetimes, one to Minkowski space as we found earlier, the other to Petrov
type N spacetimes.

Similarly, whenever the isotropy group decreases through the Karlhede
algorithm, i.e., whenever an $n^{\rm th}$ covariant derivative is used to fix
the frame, some limits might be missed. Nevertheless, as a classification for
the covariant derivatives of the Riemann tensor does not exist yet, a
different procedure will be used. In the present case, we apply the isotropy
group $H_{0}$ of $C_{0}$ given by (\ref{boost})--(\ref{spin}) to the Cartan
scalars, obtaining (\ref{boosti1a})--(\ref{boosti1b}).  From the hereditary
property, in the limit only $\widetilde{D\Psi}_{2 0^{'}}$ and
$\widetilde{D\Psi}_{3 1^{'}}$ may be non-zero. A boost (\ref{boost}) can then
be applied to make them equal up to a sign. As none of these two
transformations changes the relative sign of the components, eq.~(\ref{r1b})
still holds in the limit. So we do not really lose any Petrov type D limit
and the following theorem holds:
\begin{theorem}\label{D} The only Petrov type D or 0 local, non-singular
limits of the Schwarzschild family are:
\begin{description}
\item as $m\ti,$ the Minkowski (type 0) and the Kasner (type D) spacetimes;
\item as $m\tz,$ the Minkowski spacetime.
\end{description}
\end{theorem}

Now we shall deal with the Petrov type N limits and show that indeed they can
occur. For any Petrov type N solution a frame may be chosen such that
\cite{Karlhede}
\begin{equation} \label{cn} \begin{array}{l}
\Psi_0 = \Psi_1 = \Psi_2 = \Psi_3 = 0, \\
\Psi_4 = 1.
\end{array} \end{equation}
To transform our previous frame to one in which the Cartan scalars might have
the above limit we can use a null rotation
\begin{equation} \label{null} \left( \begin{array}{cc}
1 & 0 \\
z & 1\\
\end{array} \right), \end{equation}
with $z$ complex. In order to have a
sufficiently generic frame suitable for all Petrov type N limits, we apply the
isotropy group of $C_{0}$ in the standard form (\ref{cn}). This is again a
null rotation \label{Karlhede} and the product of these two null rotations is
also a null rotation with some complex parameter $z$. The transformed set
$C_1$ is (for sake of brevity we do not display the whole set $C_{2}$)
\begin{eqnarray}
\eb[12em][r]{$0^{\rm th}$ derivative:\hfill
$\tilde{\Psi}_2$} & = & \Psi_2, \label{t0a} \\
\eb[12em][r]{$\tilde{\Psi}_3$} & = & 3z\Psi_2, \label{t0b} \\
\eb[12em][r]{$\tilde{\Psi}_4$} & = & 6z^2\Psi_2, \label{t0c} \\
 & & \nonumber \\
\eb[12em][r]{$1^{\rm st}$ derivative:\hfill
$\widetilde{D\Psi}_{2 0^{'}}$} & = & D\Psi_{2 0^{'}}, \label{t1a} \\
\eb[12em][r]{$\widetilde{D\Psi}_{2 1^{'}}$}
& = & \bar z D\Psi_{2 0^{'}}, \label{t1b} \\
\eb[12em][r]{$\widetilde{D\Psi}_{3 0^{'}}$}
& = & 3z D\Psi_{2 0^{'}}, \label{t1c} \\
\eb[12em][r]{$\widetilde{D\Psi}_{3 1^{'}}$}
& = & (3z\bar z - 1)D\Psi_{2 0^{'}}, \label{t1d} \\
\eb[12em][r]{$\widetilde{D\Psi}_{4 0^{'}}$}
& = & 6z^2 D\Psi_{2 0^{'}}, \label{t1e} \\
\eb[12em][r]{$\widetilde{D\Psi}_{4 1^{'}}$}
& = & (6z^2\bar z - 4z)D\Psi_{2 0^{'}}, \label{t1f} \\
\eb[12em][r]{$\widetilde{D\Psi}_{5 0^{'}}$}
& = & 10z^3 D\Psi_{2 0^{'}}, \label{t1g} \\
\eb[12em][r]{$\widetilde{D\Psi}_{5 1^{'}}$}
& = & (10z^3\bar z - 10z^2)D\Psi_{2 0^{'}}, \label{t1h}
\end{eqnarray}
where $\Psi_2$ and $D\Psi_{2 0^{'}}$ are given by (\ref{r0}) and (\ref{r1a}).
Here and in what follows we use a bar to denote complex conjugation.

In order to obtain the standard form (\ref{cn}), the limits of $\Psi_2$ and
$z$ must be
\begin{equation} \label{lia} \begin{array}{ccccccc}
\Psi_2 & \rA & 0 & ~~{\rm and}~~ & z & \rA & \infty,
\end{array} \end{equation}
in such a way that the limits of the products in (\ref{t0b}) and (\ref{t0c})
are
\begin{equation} \label{lib} \begin{array}{ccccccc}
z\Psi_2 & \rA & 0 & ~~{\rm and}~~ & 6z^2\Psi_2 & \rA & 1.
\end{array} \end{equation}
These conditions hold if and only if
\begin{equation}
z^2 \rA \frac{1}{6\Psi_2}, \label{zlimit}
\end{equation}
which is a coordinate-free definition.

As singular limits are not considered in this paper, only the component with
the highest power of $z$ (the last one) among $\widetilde{D\Psi}_{A B^{'}}$
(eqs.  (\ref{t1a})--(\ref{t1c})) may not tend to zero. Its limit can be
written as
\begin{equation}
\widetilde{D\Psi}_{5 1^{'}} \rA 10z^3\bar z D\Psi_{2 0^{'}}. \label{dpsi}
\end{equation}
Substituting the expression for $D\Psi_{2 0^{'}}$ from (\ref{r1a}) into
(\ref{dpsi}) and using the limit (\ref{zlimit}) we obtain
\begin{equation}
\widetilde{D\Psi}_{5 1^{'}} \rA \frac{5}{6\sqrt{2}}\frac{\bar z}{z}
\Psi_2^{-\frac{2}{3}}\sqrt{m^{-\frac{2}{3}} + 2\Psi_2^{\frac{1}{3}}}.
\label{dl1}
\end{equation}
 From (\ref{lia}), the term outside the square root tends to $\infty$,
therefore the square root must tend to zero in order to have a bounded limit.
This implies that $m \ti$.

Similarly, applying the transformation (\ref{null}) to the second covariant
derivative (\ref{r2a})--(\ref{r2c}) we find that only the last component does
not necessarily vanish in the limit. Its limit is
\begin{equation}
\widetilde{D^2\Psi}_{6 2^{'}} \rA 15z^4\bar z^2 D^2\Psi_{2 0^{'}} +
40z^3\bar z D^2\Psi_{3 1^{'}}. \label{d2psia}
\end{equation}
Substituting (\ref{r2a}) and (\ref{r2b}) into this expression, and using the
limits (\ref{zlimit}) and (\ref{dpsi}) we have
\begin{equation}
\widetilde{D^2\Psi}_{6 2^{'}} \rA \frac{6}{5}(\widetilde{D\Psi}_{5 1^{'}})^2
- \frac{5}{3} \frac{\bar z}{z}. \label{d2psib}
\end{equation}

 From now on we shall consider $m\geq 0$. Therefore, from eqs.~(\ref{i0}) and
(\ref{i1a}) we infer that $\Psi_2$ is a real negative function, while
$D\Psi_{2 0^{'}}$ is either real positive or pure imaginary positive.  These
properties are hereditary in the sense of Geroch. Thus, from (\ref{zlimit})
the limit of $z$ is pure imaginary and from (\ref{dpsi}),
$\widetilde{D\Psi}_{5 1^{'}}$ is either real negative or pure imaginary
negative. Thus the nonzero limits of the Cartan scalars as $m\ti$ may now be
written as
\begin{eqnarray}
\eb[12em][r]{$0^{\rm th}$ derivative:\hfill
$\tilde{\Psi}_4$} & \rA & 1, \label{psin} \\
 & & \nonumber \\
\eb[12em][r]{$1^{\rm st}$ derivative:\hfill
$\widetilde{D\Psi}_{5 1^{'}}$} & \rA &  -\frac{5}{6\sqrt{2}}
\Psi_2^{-\frac{2}{3}}\sqrt{m^{-\frac{2}{3}} + 2\Psi_2^{\frac{1}{3}}},
\label{dl2} \\
& & \nonumber \\
\eb[12em][r]{$2^{\rm nd}$ derivative:\hfill
$\widetilde{D^2\Psi}_{6 2^{'}}$} & \rA &
\frac{6}{5}(\widetilde{D\Psi}_{5 1^{'}})^2 + \frac{5}{3}. \label{d2psi}
\end{eqnarray}

This set of Cartan scalars gives rise to Petrov type N metrics with a
two-dimensional $H_0$ given by (\ref{null}), $t_0 = 0$ and $d_0 = 4$. The
first covariant derivative $\widetilde{D\Psi}_{5 1^{'}}$ may take three
limits, namely: (1) constant different from zero, (2) a function of the four
coordinates, $f(x^i)$, or (3) 0. From these, the limit of
$\widetilde{D^2\Psi}_{6 2^{'}}$ may be calculated using (\ref{d2psi}). We
shall now study these three limits of $\widetilde{D\Psi}_{5 1^{'}}$.

\begin{enumerate}

\item $\widetilde{D\Psi}_{5 1^{'}} \rA {\rm constant}$. For each real or
imaginary pure negative value of this constant, satisfying (\ref{d2psi}), an
inequivalent spacetime is obtained. The isotropy group $H_0$ of $C_1$ is found
to be equal to $H_0$. There is no independent function of the coordinates, so
$d_1 = d_0 = 4$. Therefore the Karlhede algorithm stops at the first covariant
derivative. The orbit of the isometry group has dimension 4 which implies that
the metric is spacetime homogeneous. The isotropy group is two-dimensional and
the isometry group has dimension 6.

\item $\widetilde{D\Psi}_{5 1^{'}} \rA~f(x^i)$. There is one independent
function of the coordinates, namely, the function $f(x^i)$, so $t_1 = 1$.
Therefore the Karlhede algorithm proceeds to the second covariant derivative
(\ref{d2psi}) and stops with $H_2 = H_1 = H_0$ and $t_2 = t_1 = 1$. Hence, the
orbit of the isometry group has dimension 3, the isotropy group is
two-dimensional and the dimension of the isometry group is 5.

\item $\widetilde{D\Psi}_{5 1^{'}} \rA 0$. From (\ref{d2psi}),
$\widetilde{D^2\Psi}_{6 2^{'}} \rA 5/3$. This set of scalars cannot correspond
to any metric as the covariant derivative of a vanishing spinor must also
vanish.

\end{enumerate}

Instead of analysing the full integrability conditions
\cite{KarlhedeLindstron}, we searched the {\sc classi} database \cite{ISCA}
for spacetimes with those Cartan scalars. We found that those spacetimes are
special cases of the plane-wave class (21.44) in ref. \cite{KSMH}, which can be
written as
\begin{equation}
ds^{2} = [\bar A(u)z^2 + A(u)\bar z^2 + B(u)z\bar z]du^2 + 2dudv - 2dzd\bar z,
\label{pw}
\end{equation}
where $A(u)$ is complex and $B(u)$ is real. Setting $B(u)$ to zero in order
to obtain a vacuum spacetime, the nonvanishing Cartan scalars are:
\begin{eqnarray}
\eb[12em][r]{$0^{\rm th}$ derivative:\hfill
$\tilde{\Psi}_{4}$} & = & 1, \label{pwi0} \\
 & & \nonumber \\
\eb[12em][r]{$1^{\rm st}$ derivative:\hfill
$\widetilde{D\Psi}_{5 1^{'}}$} & = &
A^{-\frac{5}{4}}\overline{(A^{-\frac{1}{4}})}A_{,u}~,\label{pwi1}   \\
 & & \nonumber \\
\eb[12em][r]{$2^{\rm nd}$ derivative:\hfill
$\widetilde{D^2\Psi}_{6 2^{'}}$} & = &
A^{-\frac{3}{2}}\overline{(A^{-\frac{1}{2}})}A_{,uu}~, \label{pwi2}
\end{eqnarray}
where we have used tilde for compatibility with the notation we have been
using in the type N analysis. Here and henceforth $A_{,u}$ and $A_{,uu}$
denote the first and second derivative of $A$ with respect to the
coordinate $u$.

For specific functions $A$, the limits (1) and (2) above are
recovered. Indeed:
\begin{enumerate}
\item $\widetilde{D\Psi}_{5 1^{'}} \rA {\rm constant}$. According to
\cite{KSMH}, in the paragraph below eq. (21.44), the metric (\ref{pw}) is
homogeneous if and only if
\begin{equation}
A(u) = A_0 e^{i\kappa u} \label{pwc1}
\end{equation}
or
\begin{equation}
A(u) = A_0 u^{2i\kappa - 2}, \label{pwc2}
\end{equation}
where $A_0 > 0$ and $\kappa$ are real constants. Then, from (\ref{pwi1}) we
obtain
\begin{equation}
\widetilde{D\Psi}_{5 1^{'}} = iA_0^{-\frac{1}{2}}\kappa \label{pws1}
\end{equation}
or
\begin{equation}
\widetilde{D\Psi}_{5 1^{'}} =
-2A_0^{-\frac{1}{2}} + i2A_0^{-\frac{1}{2}}\kappa~,
\label{pws2}
\end{equation}
respectively. Choosing $\kappa = - 1$ in (\ref{pws1}) and $\kappa = 0$ in
(\ref{pws2}) all pure imaginary negative and real negative constant values of
$\widetilde{D\Psi}_{5 1^{'}}$ are recovered.

 From (\ref{pwi2}), the second covariant derivative becomes
\begin{equation}
\widetilde{D^2\Psi}_{6 2^{'}} = - \frac{\kappa^2}{A_0} \label{pw2s1}
\end{equation}
or
\begin{equation}
\widetilde{D^2\Psi}_{6 2^{'}} = - \frac{4\kappa^2}{A_0} +
\frac{6}{A_0} - i\frac{10\kappa}{A_0}. \label{pw2s2}
\end{equation}
Substituting (\ref{pws1})--(\ref{pw2s2}) into (\ref{d2psi}) we find only
two possible solutions for $A(u)$, namely
\begin{equation}
A(u) = \frac{3}{25} e^{-iu}  \label{sol2}
\end{equation}
or
\begin{equation}
A(u) = \frac{18}{25} u^{-2}. \label{sol3}
\end{equation}

\item $\widetilde{D\Psi}_{5 1^{'}} \rA~f(x^i)$. Substituting (\ref{pwi1}) and
(\ref{pwi2}) into (\ref{d2psi}), the following equation for $A(u)$ is obtained:
\begin{equation}
A_{,uu} - \frac{6}{5}A^{-1}(A_{,u})^2 -
\frac{5}{3}A^{\frac{3}{2}}\overline{(A^{\frac{1}{2}})} = 0. \label{difeq}
\end{equation}
Among the solutions of this equation, those of the form (\ref{pwc1}) and
(\ref{pwc2}) have already been discussed in the previous case. Any other
solution gives rise to the present limit case and similarly to the Petrov type
D limit (1.b), namely, $m\ti$ and $\Psi_2\rA f(x^i)$, they lead to a
single limit spacetime.

\end{enumerate}

Therefore, three Petrov type N limits as $m\ti$ where found. They are special
cases of the metric (\ref{pw}). Two are spacetime homogeneous with $A(u)$
given by~(\ref{sol2}) and~(\ref{sol3}). The third one is an inhomogeneous
spacetime where $A(u)$ is a solution of eq.~(\ref{difeq}), provided the
conditions in the paragraph below this equation are take into consideration.

No Petrov type N limit of the Schwarzschild family was found as $m\tz$.

\section{Conclusion} \label{con} \setcounter{equation}{0}

Using techniques borrowed from the equivalence problem of metrics, we have
developed a coordinate-free approach to finding limits of spacetimes. Using
this approach, we have studied the local non-singular limits of the
Schwarzschild 1-parameter family as the mass $m$ tends to zero or infinity. As
$m\tz$ only the Minkowski space (Petrov type 0) was found as the limit. As
$m\ti$ five limits were found: the Minkowski space, a Kasner spacetime of
Petrov type D, and two homogeneous and one inhomogeneous vacuum plane wave
solutions of Petrov type N\@. No other Petrov type is possible since the
Schwarzschild metric is Petrov type D.

The Petrov type D and 0 limits we found were already known \cite{Geroch}. One
of the advantages of our coordinate-free approach is that by using it we have
proved that they are the only local non-singular Petrov type D and 0 limits of
the Schwarzschild spacetime as $m\tz$ and $m\ti$ (theorem \ref{D}).  Another
important feature is that we have presented a systematic way of searching for
limits. Indeed, the three Petrov type N limits we have found are new as far as
we know. Although we feel we have covered all Petrov type N limits of the
Schwarzschild spacetime when $m\tz$ and $m\ti$, we cannot ensure that
refinements of the coordinate-free approach we have presented could not lead
to some other limit spacetime.

In his paper Geroch \cite{Geroch} discusses several hereditary properties.
One of them states that the dimension of the isometry group either increases
or remains the same in the limiting process. All the limits found here are in
full agreement with this result.

After analysing the limits of the Schwarzschild spacetime, the reasons
mentioned in section \ref{eq} for the advantages of our approach seem to be
clearer. Firstly, we are dealing with scalars so that, in a given fixed frame,
all limits can be directly determined, as we did in order to find the limits
of Petrov types 0 and D\@. Note that in coordinates, the limits are taken
using the components of the metric tensor. Secondly, the frames which may lead
to new limits are related to the initial one by subgroups of the Lorentz
group, while coordinate bases transform according to GL(4,R).

Although the presentation of our approach has been mainly in terms of the
Schwarzschild spacetime, which is a Petrov type D vacuum solution of
Einstein's field equations, the techniques we have introduced can be extended
to other situations. For non-vacuum spacetimes, e.g., one ought to take into
account the Segre classification. To analyse limits of metrics with other
Petrov types, frame transformations different from (\ref{null}) have to be
considered.

Two improvements on the equivalence techniques would certainly bring
refinements to our approach: firstly, the classification of the covariant
derivatives of the Riemann tensor, analogous to the Petrov and Segre
classifications; secondly, a full study of the integrability conditions on the
Cartan scalars \cite{KarlhedeLindstron}.

We have made use throughout this paper of one of Geroch's hereditary
properties that we discussed in section \ref{eq}. Nevertheless, since the
Cartan scalars fully describe the spacetime, we believe that even the
hereditary properties of limits may be proved using the techniques outlined
here. Another future development would be an algorithm, or similar, to find
the coordinate systems which give the predicted limits, or at least a proof of
the existence of such systems. Moreover, it would be interesting to include in
this analysis singular limits, especially those which are singular only on a
hypersurface.  This would extend our analysis of the Schwarzschild spacetime
to include the singular limits presented in~\cite{AichelburgSexl},
\cite{DrayHoof} and~\cite{FerrariPendenza}.

\section*{Acknowledgments}

F. M. Paiva would like to thank his colleagues at Queen Mary and Westfield
College where part of this work was done, the Conselho Nacional de
Desenvolvimento Cient\'{\i}fico e Tecnol\'ogico (CNPq) for financial support
and Jim Skea for his help with the {\sc classi} database.  F. M. Paiva and M.
J. Rebou\c{c}as thank A. F. F. Teixeira and F. J. Wright for useful comments.

\end{document}